# Translucency and negative temperature-dependence for the slip length of water on graphene

Han Li[1,2], Zhi Xu[1,2], Chen Ma[2,3], Ming Ma[1,2]*

[1]Department of Mechanical Engineering, State Key Laboratory of Tribology, Tsinghua University, Beijing 100084, China

[2]Center for Nano and Micro Mechanics, Tsinghua University, Beijing 100084, China

[3]Department of Engineering Mechanics, Tsinghua University, Beijing 100084, China

**Abstract**

Carbonous materials, such as graphene and carbon nanotube, have attracted tremendous attention in the fields of nanofluidics due to the slip at the interface between solid and liquid. The dependence of slip length for water on the types of supporting substrates and thickness of carbonous layer, which is critical for applications such as sustainable cooling of electronic devices, remains unknown. In this paper, using colloidal probe atomic force microscope, we measured the slip length of water on graphene $l_s$ supported by hydrophilic and hydrophobic substrates, *i.e.*, $SiO_2$ and octadecyltrimethoxysilane (OTS). The $l_s$ on single-layer graphene supported by $SiO_2$ is found to be 1.6±1.9 nm, and by OTS is 8.5±0.9 nm. With the thickness of few-layer graphene increases to 3~4 layers, both $l_s$ gradually converge to the value of graphite (4.3±3.5 nm). Such thickness dependence is termed slip length translucency. Further, $l_s$ is found to decrease by about 70% with the temperature increases from 300 K to 350 K for 2-layer graphene supported by $SiO_2$. These observations are explained by analysis based on Green-Kubo relation and McLachlan theory. Our results provide the first set of reference values for the slip length of water on supported few-layer graphene. They can not only serve as a direct experimental reference for solid-liquid interaction, but also provide guideline for the design of nanofluidics-based devices, for example the thermo-mechanical nanofluidic devices.

Keywords: water, graphene, slip length, atomic force microscope, nanofludics

# Introduction

The no-slip boundary condition is one of the most influential hypotheses in fluid mechanics, assuming that the velocity of liquid equals to the velocity of solid at the solid-liquid interface. However, in molecular scale, liquid molecules may slip on the solid surfaces, leading to different velocities of liquid and solid at the interface[1, 2]. Such phenomenon is called boundary slip and is quantified by "slip length" ($l_s$). The definition of slip length in continuum fluid mechanics is $v_l/l_s = (\partial v_l/\partial z)|_{z=0}$, given by Navier[3], where $z$ is the normal coordinate and $v_l$ is the lateral velocity of liquid parallel to the solid surface (Figure 1a)[1, 2]. On molecular scale, the slip length is defined as $l_s = \eta/\lambda$, where $\eta$ is the liquid viscosity and $\lambda$ is the interfacial friction coefficient between liquid molecules and solid surface[4]. Both definitions rely on the continuum assumption of fluid[1, 4]. According to N. Kavokine and L. Bocquet et al, for water, if the characteristic scale of the fluid field is larger than 1 nm, the flow can be regarded as continuum and the classical Navier-Stokes Equation works, which indicates the applicable scope of slip length[4].

Since the 1823 memoire by Navier[3], there has been a great century of science about the slip length, and even more, where boundary slip was replaced with no slip because experimental apparatus was not precise enough to account for slip. Since 1950s, based on the booming developments of experimental techniques such as microfluidic, surface apparatus and atomic force microscope, people finally could see slippage[2, 4-7]. It has been proved that several factors may affect $l_s$, such as the wettability[8], roughness[2], viscosity[9], degree of confinement[10], velocity[11] et al. Most slips occur on hydrophobic or semi-hydrophobic surfaces. In essence, $l_s$ is closely correlated with the interfacial intermolecular and surface forces, and ranges from several nanometers[10, 12, 13] to tens of micrometers[2]. As a result, the boundary slip behavior mainly affects micro- and nanoscale flow field and plays important roles in both scientific and engineering fields, such as interfacial mechanics[7], lubrication[14], bearing design[15], surface drag reduction[16], liquid transport[17], micro- and nanofluidics[12, 18], micro- and nano-electro-mechanical systems (MEMS/NEMS)[6] et al. For example, it has been reported that in both macro- and micro- scale, compared with traditional no-slip surfaces, the journal bearings with large slip surfaces have higher load capacity and lower energy cost[6, 15].

Recently, an inspiring experiment showed that the micro-/nanofluidics could be applied as a sustainable cooling method in electronic devices[19]. The heat fluxes of the mentioned microfluidic cooling design exceed 1.7 kW/cm$^2$, with an unprecedented coefficient of performance about $10^4$, 10 ~100 times better compared with traditional design[19]. For such micro-/nanofluidic devices, the heat fluxes are determined by both the fluid transport and energy transport properties[20]. Obviously, the slip length which determines the transport rate has a great influence in the cooling performance.

Because of the atomic smoothness and thickness, two dimensional (2D) materials have attracted increasing attention in the field concerning solid-liquid interface. The extensive experiments on carbonous nanofluidics, such as carbon nanotubes (CNTs)[10, 13], graphite/graphene nanochannels[12, 21, 22] prove that graphite/graphene could significantly increase the transport rate of water in nanochannels by increasing the slip length (50~60 nm for graphite[20-21], and 16±100 nm for graphene supported by silica[12]) on the promise of keeping an smooth and accurate shape of the channels. Nanoscale solid surface curvature[23] and confinement[1, 4], types of supporting substrate[24, 25] are believed to have significant influence on the slip length of carbonous materials: (1) the curvature will induce incommensurability between water and carbon atomic structure, which increase the slip length[23]; (2) the substrate may affect the adhesion energy and energy barrier and modulate the slip length[24, 25]; (3) the nanoscale confinement leads to a prominent interfacial effect where the density and viscosity in the fluid field vary, thus affects the slip length[1, 4, 26]. Recently, G. Greenwood and M. Espinosa-Marzal et al measured the slip length of sucrose, silicone oil and ionic liquids on supported single-layer graphene and found that the slip length strongly depends on the types of supporting substrates and liquid[25].

For the application of fluids under micro- or nano-scale confinement, such as the above-mentioned journal bearings and microfluidic cooling devices, not only the flow resistance at the solid-liquid interface plays an important role, but also the thermal resistance.[20] The former is characterized by slip length, and the latter is characterized by the thermal conductivity of the interface, including Kapitza resistance and the thermal conductivity of the coatings. From this aspect, graphene is a good candidate due to its finite slip length with water[12] and high thermal conductivity[27, 28]. Series of works have been focused on the effects of the thickness of graphene layer[27, 28] and the types of supporting substrates[29] on the thermal conductivity, revealing rich phenomena. However, their effects on slip length remains a mystery, casting a shadow for the application of carbonous materials in nanofluidic devices.

In this paper, the slip length of water on graphene, $l_s$, supported by silica ($SiO_2$) or octadecyltrimethoxysilane (OTS) substrates is measured. Measurements on bare $SiO_2$ and OTS surfaces are also performed for comparison. Our results show that for single-layer graphene (SLG) on $SiO_2$ (SLG/$SiO_2$) $l_s$ is 1.6±1.9 nm, close to no-slip, while for SLG on OTS (SLG/OTS) $l_s$ is 8.5±0.9 nm. The slip length of water on bare $SiO_2$ and OTS substrates in our experiments is 0.9±1.2 nm and 13.4±3.0 nm respectively. With the thickness of few-layer graphene (FLG) increasing, $l_s$ converges at about 3~4 layers thick. We also measured $l_s$ of a relatively flat bulk graphite whose value is found to be 4.3±3.5 nm, close to the reported value in literature (4.5±4.4 nm[30] and 8 nm[31, 32]) and the convergent value of FLG as measured here (4.6±1.7 nm). With careful analysis, we excluded the potential effects of surface roughness on the variation of $l_s$. Thus, the dependence of slip length on the thickness of FLG reflects the partially shielding effects of graphene on the interaction between water and the substrate ($SiO_2$ or OTS). Such a phenomenon is termed as *slip length translucency* here, similar to the previous reported wetting translucency[33]. Further measurements on temperature-dependence for $l_s$ reveal that the comprehensive effect of temperature, water density, viscosity, dielectric constant, and force relaxation time play a vital role in slip.

**Results**

The slip length $l_s$ was measured by the colloidal probe atomic force microscope (AFM, Cypher ES) as shown in Figure 1a, which is commonly used in the hydrodynamic force and surface force measurements.[34] The temperature was controlled by the hot-and-cooling plate mounted on the sample stage of AFM. To avoid thermal drifts, only when the AFM cantilever deflection signal was stable at given preset temperature, the measurement was performed. The samples used in this paper were characterized by Raman spectrum and AFM to get the layer number of graphene and the surface roughness (Figures 1b and 1c). The atomic images of graphene sample (Figure 1d and Figure S6 in supplementary information) indicate the smoothness and intact structure of graphene. Before each measurement, the graphene samples were annealed at 120°C for 20 minutes in air to avoid the adsorbed air-born contamination[35]. The colloidal probe consists of a tipless cantilever (CSC38, Mikromasch

Instrument) and an ultraviolet curably glued microsphere tip (borosilicate glass, 9020, Duke Scientific). The hydrophilic microsphere of which the slip length is zero was used to amplify the force signal acting on the probe by the surrounding water. The radius of the microsphere $R$ is 10 μm, and the root-mean-square (RMS) roughness is about 1.2 nm for a 400×400 nm$^2$ area on the microsphere, which is as clean and smooth as previous literatures[2, 36]. Degassed deionized water (18.2 MΩ, Hitech Sciencetool) was used for slip length measurement. During the measurement, the sample was controlled by a piezoelectric ceramic stage to approach to and retract from the probe at a given speed $v_{\text{piezo}}$, and the hydrodynamic force acting on the tip was recorded by the photo detector as voltage signal, which can be further processed into force.[34] The force depends sensitively on the slip length, reflected by the force-distance curve[37]. For such setup, O. I. Vinogradova, D. F. Honig, W. Ducker, and C. Cottin-Bizonne et al, provide the formula to estimate the slip length[30, 37, 38], $v/F_{\text{h}} = (h + l_{\text{s}})/6\pi\eta R^2$, where $h$ is the separation between the microsphere and the surface of the substrate as shown in Figure 1a, $v = \dot{h}$ is the velocity of the sphere surface, $F_{\text{h}}$ is the hydrodynamic force acting on the tip, $\eta$ is the viscosity of liquid. Therefore, the slip length can be fitted from the hydrodynamic force acting on the tip (Figure 1e). In reference to previous slip length experiments[30, 38], the curve fitting range for our measurement is $h \in (20, 200)$ nm where the hydrodynamic force is significant and sensitive to the slippage in our experimental setup (Figure 1e). Standard deviation with multiple measurements was chosen as the error bar. More details of the measurements are given in supplementary information (SI). For our present setup, a detailed analysis considering various error factors and error transfer was applied. The systematic factors contain the thermal drifts, virtual deflection, liquid viscosity, and cantilever stiffness error. The random factors contain thermal noise, zero force point, contact point, roughness of microsphere and inverse optical lever sensitivity. The analysis results show that after the error transfer, the overall measurement accuracy (standard deviation) for $l_{\text{s}}$ is 0.7 nm (see SI for more details). The contact angle $\theta$ of SiO$_2$ and OTS was measured to be 24.9° and 100.0° respectively, confirming the hydrophilic/hydrophobic nature of the substrates (see SI for details).

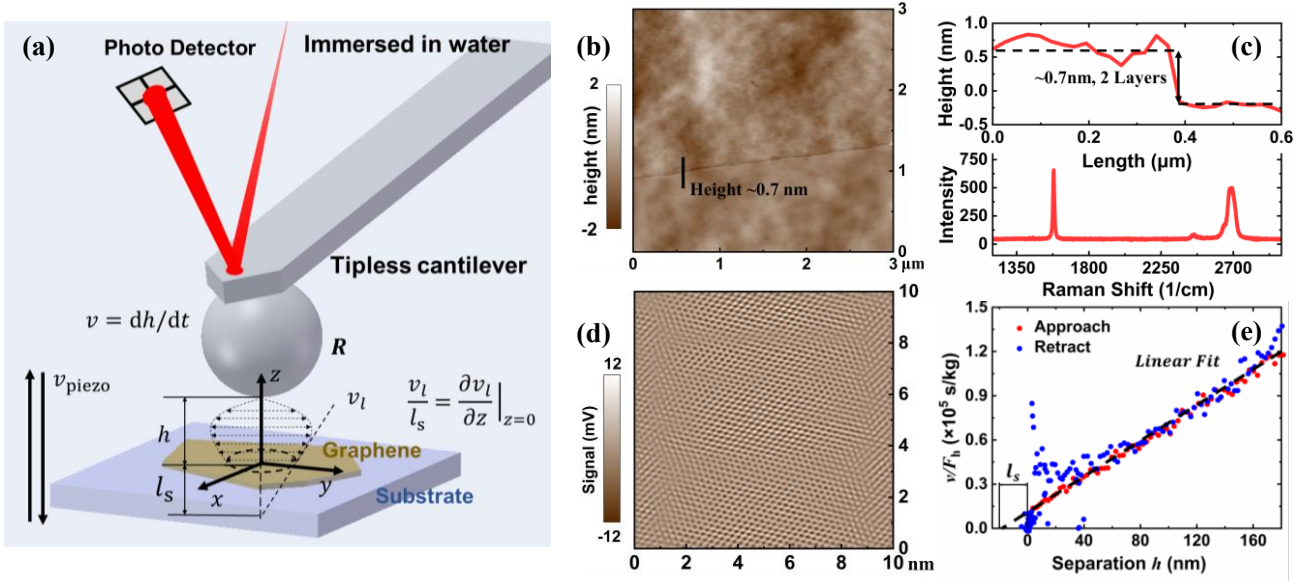

**Figure 1. Experimental setup and measured force curves.** (a) Schematic of colloidal probe atomic force microscope. (b) AFM morphology around the edge step of a 2-layer graphene supported on $SiO_2$. (c) Step height (upper) and Raman shifts (lower) of 2-layer graphene, the step is measured along the black line in (b). (d) Typical atomic image of the graphene layer characterized by the lateral force signal (filtered by fast Fourier Transform). (e) $v/F_h$-separation curve.

For bare substrates, on $SiO_2$ the measured $l_s$ is 0.9±1.2 nm, indicating a no-slip boundary condition[5, 30]. On OTS $l_s$ is 13.4±3 nm, corresponding to the slip boundary condition, which agrees with previous reports (10~30 nm, typically)[38]. On highly ordered pyrolytic graphite (HOPG), because of the steps and waviness, different locations may lead to different slip length values [2]. Here we chose a relatively flat location with RMS about 2 nm for a 20×20 μm² area and found $l_s$ to be 4.3±3.5 nm. This is close to the value reported by C. D. F. Honig and W. A. Ducker (4.5±4.4 nm)[30], and the value reported by A. Maali, Y. L. Pan and B. Bhushan et al (8 nm)[31, 32]. The consistency between our values and those reported for graphite confirms the validity of our measurements.

For supported FLG, our measurements began with the velocity dependence as it is one of the most widely studied properties of slip length. For each value of $v_{piezo}$, being 75, 105, 150 μm/s respectively, we measured $l_s$ over tens of times. It should be noted that $v_{piezo}$ is different from but related to the actual velocity of the probe $v$[34]. The values of $l_s$ for different $v_{piezo}$ are listed in Figure 2b and Table S1 in SI, showing that $l_s$ is almost independent on $v_{piezo}$.

In previous studies, there are two opposite views about the influence of velocity on slip length[5]. Most experiments support that $l_s$ of simple liquids is independent of $v_{piezo}$[2, 36], which agree with our present observation. These are because the typical value of $v_{piezo}$ in AFM experiments (< 200 μm/s, corresponding to shear rate $10^3 \sim 10^4$ s$^{-1}$)[36] is much smaller than the critical velocity that may lead to a slip variation ($10^6 \sim 10^9$ s$^{-1}$, typically)[39, 40]. Aiming at revealing the mechanism of slip, to focus on the influence of other factors, as $l_s$ is practically independent of velocity in our study, we averaged $l_s$ for different $v_{piezo}$ to get better statistics. For convenience, the symbol $l_s$ is kept denoting the averaged slip length afterwards.

The feasibility to control the thickness of FLG with sub nanometer precision enable us to fine-tune the interaction between liquid and solid on atomic level to study the slip mechanism. Therefore, we measured the dependence of $l_s$ on the thickness of FLG. As shown in Figure 2a, for single layer graphene, $l_s$ on SLG/SiO$_2$ is 1.6±1.9 nm while on SLG/OTS it is 8.5±0.9 nm, reflecting the hydrophilic and hydrophobic nature of the two bare substrates ($l_s$ being 0.9±1.2 nm for bare SiO$_2$, and 13.4±3 nm for bare OTS). However, $l_s$ of 2-layer FLG on both substrates are closer to each other, being 3.6±1.6 nm and 5.7±2.5 nm respectively. For the 3, 4, and 6-layer FLG on SiO$_2$, $l_s$ is 4.6±1.4, 4.4±2.7 and 4.6±1.7 nm respectively, gradually converged to the value on HOPG we measured (4.3±3.5 nm). Similarly, for the 5 and 9-layer FLG samples on OTS, the values are 4.0±1.9 and 4.2±1.3 nm, also close to the results measured on HOPG (4.3±3.5 nm). Because the mechanical exfoliation of FLG on OTS is much more difficult than that on SiO$_2$, we didn't perform the measurement of 3 and 4-layer FLG on OTS. However, it is clear that for $l_s$ of FLG on different substrates, the values of $l_s$ on single-layer graphene are very different and close to that of the substrate, and the values on FLG with number of graphene layers $N$ larger than 3 are the same as those measured on smooth HOPG, with $l_s$ for $N = 2$ in between. Such a thickness dependent slip length phenomenon is termed as *slip length translucency*.

We noticed that a similar phenomenon is the wetting transparency or translucency on 2D materials[33, 41, 42]. With the close relation between $l_s$ and $\theta$, which is called scaling law [1, 43-45], $l_s \sim (1+\cos\theta)^{-2}$, the slip length translucency shows qualitative agreement with wetting translucency. However, such an agreement becomes questionable when taking temperature into consideration. The existing experimentally measured water contact angle is insensitive to temperature below the boiling point for most surfaces[46] including graphite[47]. In contrast, the slip length is predicted theoretically temperature-dependent in the same temperature range[48, 49]. Therefore, it would be interesting to find out experimentally whether $l_s$ is temperature-dependent. Here we directly measured the dependence of $l_s$ on temperature for 2-layer graphene on $SiO_2$. The results (Figure 2b) show that $l_s$ decreases by approximately 70% (from 3.7±1.6 nm to 1.1±1.9 nm on average) as the temperature increases from 300 K to 350 K, indicating a clear temperature dependence of $l_s$.

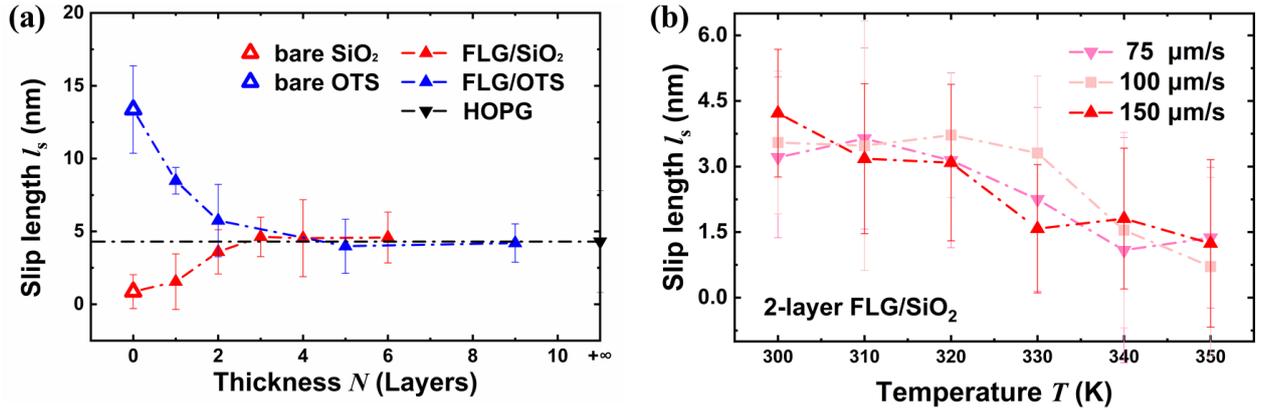

**Figure 2. Slip length measurement results.** (a) Slip length of water on graphene with different number of layers $N$ on different substrates (b) Temperature-dependent slip length of water on 2-layer graphene supported by $SiO_2$ at different velocities $v_{\text{piezo}}$.

We notice that a possible critical factor for our measurements is the surface roughness. With the thickness increasing, the morphology of supported graphene may evolute from the morphology of the substrate to the graphite. In view of the close relationship between slip length and surface roughness[2, 8, 50], a natural conjecture about mechanism of the slip length translucency is that the variation of slip length may be due to the change of solid surface roughness. However, as discussed afterwards, this conjecture can be excluded by examining the morphology of FLG samples carefully. The raw morphological images of 1, 2, 3-layer FLG on SiO$_2$ are shown in Figures 3 a~c. Qualitatively, it is evident that the morphology hardly changes with the thickness. To achieve a quantitative estimation, we further calculated the corresponding root-mean-square roughness (RMS) and peak-to-valley roughness (PV). From Figure 2a, for FLG/SiO$_2$, $l_s$ first increases sharply from $N=1$ to $N=3$ then remains a constant. The corresponding RMS roughness (Figure 3d, lower panel), however, keeps a constant with $N=1, 2$ and then oscillates, showing distinct features. Therefore, it is reasonable to conclude that the variation of $l_s$ is not due to the RMS roughness. Same conclusion could also be made for FLG/OTS. The $l_s$ decreases nonlinearly with $N$ while RMS roughness remains unchanged for $N=1,2$ and then decreases slightly by 0.08 nm from $N=2$ to $N=9$. By observing the dependence of $l_s$ and PV roughness on $N$, similarly, one could also reach the conclusion that there is no correlation between the change of $l_s$ and PV roughness in our experiment. Such observations do not conflict with previous experimental studies showing roughness-dependent $l_s$ as their roughness (RMS 10~100 nm[51-53]) is orders larger than ours (RMS < 1 nm).

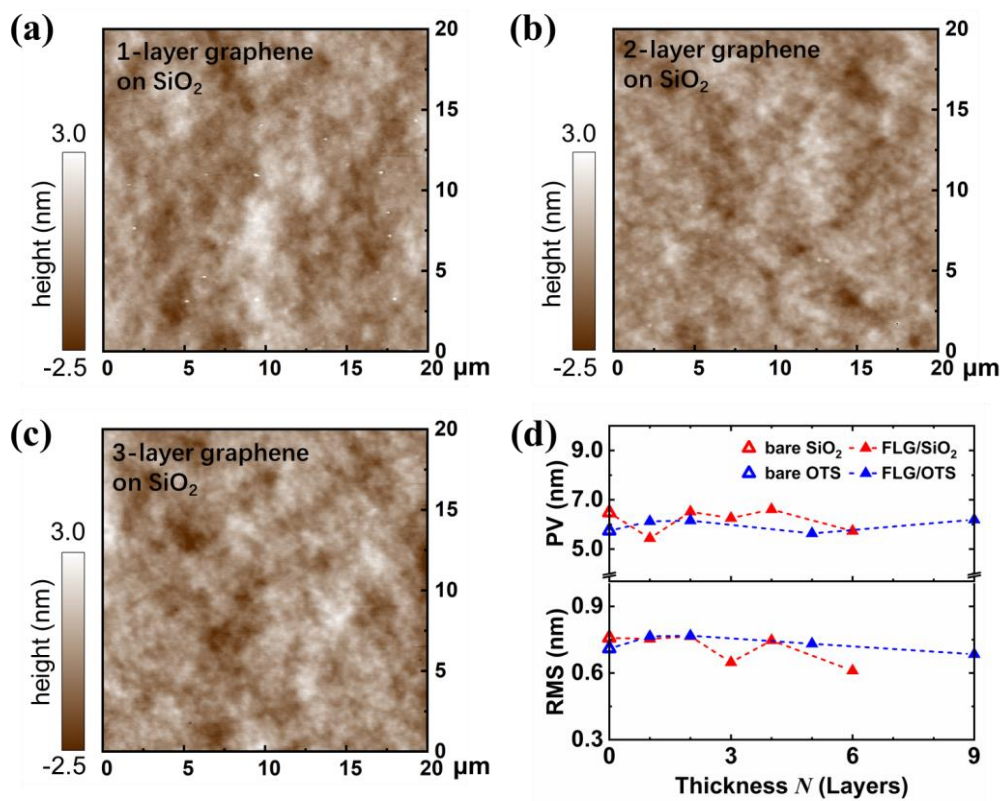

**Figure 3. Morphology of few-layer graphene.** (a~c) AFM morphological images of 1, 2, 3-layer graphene supported by SiO$_2$, respectively. (d) Root-mean-square (RMS) and peak-to-valley (PV)

roughness of few-layer graphene supported by SiO₂ or OTS.

After excluding the direct effects of roughness on the variation of $l_s$, the only possible influence of roughness is the inaccuracy of $l_s$ due to the estimation of separation $h$.[2] This is also negligible with the analysis shown afterwards. For the roughness of spherical tip, its corresponding inaccuracy can be eliminated by noticing that its effect on separation is linear[2, 38] (See SI for more details). As a result, before each measurement on FLG, we first measured the apparent slip length on mica whose true slip length is zero, then we subtracted this value from the apparent slip length on FLG to exclude the effect of the tip on $h$. For the roughness of FLG, when its characteristic wavelength is smaller than $\sqrt{2Rh}$, its impact on the separation can be regarded as shifted boundary[2], *i.e.*, the separation may be increased by a value close to the roughness. When its characteristic wavelength is larger than $\sqrt{2Rh}$, its impact on the separation is negligible[2]. For our experiments, the characteristic wavelength is about 7~10 μm (see SI for more details), which is much larger than $\sqrt{2Rh} = 2$ μm with $R = 10$ μm and $h = 0.2$ μm. Thus, the roughness of FLG also has negligible effects on the measurement of slip length.

## Discussion

Our measurements show that $l_s$ is independent of velocity within the measured range. The slip length, however, does depend on the wettability of supporting substrates and the thickness of FLG, and converges to 4 ~ 5 nm. It can be further decreased significantly by increasing the temperature. While existing reports[2, 36] could explain the velocity independence, there is no investigation for the thickness dependence, and the existing theoretical results could not explain our measured temperature-dependent slip length as discussed later.

For the mechanism for slip length translucency, we resorted to existing results on wetting translucency[1, 2]. Inspired by the continuum mechanics-based calculation about wetting translucency by C-J. Shih et al[41], we first calculated water-solid adhesion energy $W_{ad}$ for a *N*-layer-graphene-coated substrate with approach similar to C-J. Shih et al [41] using the 12-6 Lennard-Jones potential (see SI for more details). Then in view of the close relation between $l_s$ and $\theta$,[8] we calculated the estimated slip length according to $l_s \sim (1 + \cos\theta)^{-2} \sim W_{ad}^{-2} = \kappa/W_{ad}^2$ with $\kappa$ being the scaling factor[1]. We estimated $\kappa$ by fitting $l_s$ to $W_{ad}$ in the convergent regime ($N > 4$, as shown in Figure 4a). With the estimated value of $\kappa$, for $N \leq 3$, we compared $l_s$ with $\kappa/W_{ad}^2$ directly. The good agreement between $l_s$ measured experimentally and predicted theoretically ($\kappa/W_{ad}^2$) not only clearly shows the validity of the experiment (Figure 4a), but also indicates that the mechanism of slip length translucency shares common features with that of the wetting translucency.

Regarding the negative temperature dependence of $l_s$, Keliu Wu et al[49] show that only when the temperature $T$ is close to the critical temperature (~647 K for water) where the structure and dynamics of nano-confined interfacial water changes dramatically, the slip length drastically changes. Obviously, this is not consistent with our results where $T$ is < 350 K. In another report[48], with molecular dynamics (MD) simulations C. Herrero et al considered the dynamic effect of relaxation time and estimated that the slip length of water could decrease by ~20% for graphene (from ~50 nm to ~40 nm) with $T$ increasing from 300 K to 350 K. This is significantly different with our present results where $l_s$ decreases by about 70% (from 3.7±1.6 nm to 1.1±1.9 nm averagely) in the same temperature range. One possible reason for such discrepancy is that the MD simulation[48] neglects the notable dependence of intermolecular interactions on temperature, which may lead to the underestimation of the temperature effect on the slip length[7, 54]. Therefore, we will provide our preliminary explanation.

Concerning that slip length is also affected by temperature besides the adhesion energy, the scaling law between $l_s$ and $\theta$ which could explain the slip length translucency no longer works when temperature changes because $\theta$ is independent of $T$ [46, 47]. Recently, some theoretical works showed that when the solid surface charge distribution changes[55, 56], $l_s$ may vary even if $\theta$ keeps constant. Using $\Delta E$ to characterize the energy barrier of water molecule moving from one equilibrium location to its neighbor on the solid surface, the different dependences of $l_s$ and $\theta$ on surface charge distribution are attributed to the key role of $\Delta E$ on the slip length variation[55, 56]. Inspired by these studies[55, 56], we explored the temperature-dependent slip mechanism from the perspective of $\Delta E$ afterwards.

Based on Green-Kubo relations, the effect of $\Delta E$ on $l_s$ can be expressed as follows: [23]

$$\begin{cases} l_s = \eta/\lambda \\ \lambda \cong \dfrac{t_0}{k_B T}\rho_1[S_1(\boldsymbol{q}_+) + S_1(\boldsymbol{q}_-)](q_0 \Delta E)^2 \end{cases} \quad (1)$$

where $t_0$ is the force autocorrelation time, $k_B$ is the Boltzmann constant, $\rho_1$ is the density of the first water layer, $S_1$ is the two-dimensional structure factor of the first water layer, $\boldsymbol{q}_\pm$ are the reciprocal lattice vectors of graphene: $\boldsymbol{q}_\pm = q_0(1/\sqrt{3}, \pm 1)$, $q_0 \cong 25.5$ nm$^{-1}$. As we are interested in the dependence of $l_s$ on $T$, here we focus on the dependence of each variable in equations (1) on $T$. The dependence of $t_0$ and $\rho_1$ on $T$ is calculated via MD and Boltzmann distribution respectively. The structure factor $S_1$ is proved independent with $T$ by C. Herrero and L. Joly et al[48]. For $\Delta E$, we used the McLachlan theory of van der Waals force to estimate its temperature-dependence via the intermolecular interaction $w_{\text{VDW}}(r)$ between water molecule and each carbon atom:[7, 54]

$$\Delta E \sim w_{\text{VDW}}(r) = -\frac{6k_BT}{4\pi\epsilon_0 r^6}\left(\frac{1}{2}\frac{\alpha_w(iv_0)\alpha_c(iv_0)}{\epsilon_w^2(iv_0,T)} + \sum_{n=1}^{\infty}\frac{\alpha_w(iv_n)\alpha_c(iv_n)}{\epsilon_w^2(iv_n,T)}\right) \quad (2)$$

where $r$ is the distance between the water molecule and solid atom, $\epsilon_0$ is the dielectric constant of vacuum, $\epsilon_w$ is the relative dielectric constant of the water medium, $\alpha_w(iv_n)$ and $\alpha_c(iv_n)$ are the polarizabilities of water molecule and carbon atom, $iv_n$ is the imaginary frequency of molecules with $i$ as the imaginary unit ($i^2 = -1$), $n$ is the order of imaginary frequency being 0, 1, 2······. For simple polar molecules, $v_n = n(k_BT/\hbar) \cong 4n \times 10^{13}$ s$^{-1}$ at 300 K, $\hbar$ is the Planck constant. For our system, $\alpha_w$ and $\alpha_c$ are approximately independent of $T$.[7] The equation (2) can be simplified into $\Delta E \sim w_{\text{VDW}}(r) \sim T[1/2\epsilon_w^2(iv_0,T) + \sum_{n=1}^{\infty} 1/\epsilon_w^2(iv_n,T)]$.

Combine equations (1) and (2), the result is (details are shown in SI):

$$l_s = \frac{\eta}{\lambda} = \kappa_T \frac{\eta \epsilon_w^4}{\rho_1 t_0 T} \quad (3)$$

where $\eta$, $\epsilon_w$, $\rho_1$, $t_0$ are all functions of $T$, $\kappa_T$ is a scaling factor containing $\alpha_j(iv_n)$, $k_B$, $\epsilon_0$, $S(\boldsymbol{q}_\pm)$ which are all independent with $T$. Thus $\kappa_T$ is independent with $T$ within the known range of parameters studied here. Therefore, equation (3) can be used to estimate the temperature dependence of $l_s$. The fitting result of equation (3) is shown in Figure 4b, appearing good agreement with experimental results. Thus, it is reasonable to conclude that our theory could explain the experimental measurements reasonably well. We also used $\Delta E$ to analyze the slip length translucency and found agreement for all substrates (see in SI).

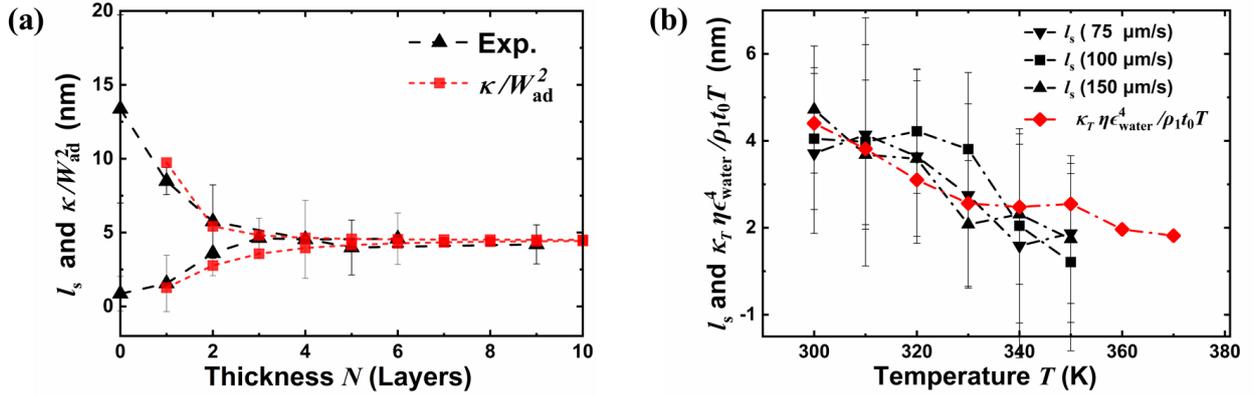

**Figure 4. Theoretical explanation of slip length translucency and temperature dependence.** (a) The agreement between $l_s$ measured experimentally and estimated theoretically with adhesion energy $W_{ad}$ for FLG in different thickness. The fitted value for $\kappa$ is $4.2\times10^{-12}$ J/m. (b) For 2-layer graphene on $SiO_2$, the comparison for $l_s$ measured experimentally and estimated theoretically based on equation (3) at different temperature with different velocities $v_{piezo}$. The fitted value for $\kappa_T$ is $5.6\times10^{-19}$ $V^4m^3s^2K/C^4$.

Together with the theoretical explanation, our measurements could provide useful guidelines for the application of fluids under micro- or nano-scale confinement. As mentioned in the introduction, not only the flow resistance at the solid-liquid interface plays an important role, but also the thermal resistance. For graphene and other 2D materials, the effects of supporting substrate and thickness of 2D layer on thermal conductance have been studied widely [27-29], however, their effects on the slip length of water still lack answers. This blank now is filled by our present slip length measurements. Together with the temperature-dependence, our results provide direct experimental reference for the designs of the upcoming 2D materials-based micro- and nanofluidic devices.

Our results could also provide rich insights in the mechanism of slip length. Together with the slip length of water in CNTs ($10\sim10^5$ nm[13, 57]), the small $l_s$ on graphene as measured here (< 10 nm) validate the current understanding, *i.e.* the curvature will induce incommensurability between water and carbon structure, leading to larger slip length[23]. The values of $l_s$ in the transition regime where graphene thickness < 1 nm could serve as a reliable benchmark for the theory of solid-liquid interaction on atomic scale. Regarding temperature dependence, several MD simulations support that $l_s$ is positively correlated with the temperature for hydrophilic surfaces (contact angle < 90°)[58, 59], while some other groups hold exactly the opposite view[48, 49]. The observed negative temperature-dependence of slip length provides the first experimental reference for such controversy[48, 49, 58, 59].

An interesting observation is that the slip length of water on graphite (4.3±3.5 nm measured here, 4.5±4.4 nm[30] and 8 nm[31]) and FLG (4.6±1.7 nm measured here) measured by AFM is much smaller than in graphite nanochannels (height = 1~30 nm, $l_s$ = 50~60 nm[20-21]). Similarly, $l_s$ on SLG/SiO$_2$ (1.6±1.9 nm) is also much smaller than in graphene nanochannels (height ~ 50 nm, $l_s$ ~16 nm[12]). Considering similar surface roughness (0.7 nm for FLG and 0.47~0.98 nm for graphene nanochannels[12]), this could be caused by the confinement of the nanochannels. However, the range where confinement plays a role in slip length (~50 nm) seems to be larger than previously thought (5 nm[21]).

## Conclusion

In summary, the slip length of water on few-layer graphene supported by SiO$_2$ or OTS was measured. The slip length translucency phenomenon was observed. The slip length was found to be velocity independent within the measured range but decreases by about 70% as the temperature increases from 300 K to 350 K. The slip length translucency is due to the variation of adhesion energy. The temperature dependence is caused by the comprehensive effect of temperature, and the dependence of water density, viscosity, dielectric constant, and force relaxation time on temperature. Our results provide the first set of reference values for the slip length of water on supported few-layer graphene. They show insightful understandings on the origin of slip and provide a guideline for the modulation of friction between water and graphene. Further studies could be guided towards the measurement of slip length of water on other 2D materials or suspended graphene.

**Acknowledgements**

M.M. acknowledges the financial support from the NSFC (Grant no. 11890673, 11772168 and 51961145304), Shenzhen Science and Technology Innovation Committee (Grant no. 2020N036) and the support from supercomputer Tansuo 100 of Tsinghua University.